\documentclass[oldversion,rnote]{aa}
\usepackage{amsmath,graphics,amssymb}

\usepackage{times}
\usepackage{natbib}
\bibpunct{(}{)}{,}{a}{}{,}

\newcommand{\zav}[1]{\left(#1\right)}

\newcommand{\kms}{\ensuremath{\mathrm{km}\,\mathrm{s}^{-1}}}
 
\newcommand{\vel}{{v}}
\newcommand{\vinfty}{\ensuremath{\vel_\infty}}
\newcommand{\vinfbet}{\ensuremath{\vel_{\infty,\beta}}}
\newcommand{\vmod}{\ensuremath{v^\text{mod}}}
\newcommand{\vapp}{\ensuremath{v^\text{app}}}

\allowdisplaybreaks

\begin{document}

\title{Improved velocity law parameterization for hot star winds}

\author{J.  Krti\v{c}ka\inst{1} \and J. Kub\'at\inst{2}}
\authorrunning{J. Krti\v{c}ka and J. Kub\'at}

\institute{\'Ustav teoretick\'e fyziky a astrofyziky P\v{r}F MU,
            CZ-611 37 Brno, Czech Republic, \email{krticka@physics.muni.cz}
           \and
           Astronomick\'y \'ustav, Akademie v\v{e}d \v{C}esk\'e
           republiky, CZ-251 65 Ond\v{r}ejov, Czech Republic}

\date{}

\abstract{The velocity law of hot star winds is usually parameterized via the
so-called beta velocity law. Although this parameterization stems from
theoretical considerations, it is not the most accurate description of the wind
velocity law that follows from hydrodynamical calculations. We show that the
velocity profile of our hydrodynamical wind models is described much better by
polynomial approximation. This approximation provides a better fit than the beta
velocity law already for the same number of free parameters.

}

\maketitle

Hot stars lose an important part of their mass via line-driven winds.
Consequently, mass-loss rate prescriptions are an important part of evolutionary
codes. Such prescriptions can be derived from theoretical modelling of hot star
winds. However, any theoretical prescription should be tested against the mass-loss
rates derived from observations. Although there are several possibilities for
deriving wind mass-loss rates (e.g., from H$\alpha$ line profiles, from
ultraviolet P~Cygni line profiles, and from radio continua), knowing of
the wind velocity law is needed to derive the mass-loss rates from
observations. Although there exist sophisticated hydrodynamic calculations
giving the wind velocity law in a tabular form \citep[e.g.,][]{graham,cmf1},
analytic formulae 
are
useful
in
many analyses. Consequently, the question
of proper velocity law parameterization is important for
analysing
observed
hot star wind spectra. Here we develop a parameterization that is as close as
possible to the results of hydrodynamical calculations.

The most frequently used parameterization of the wind velocity law
is the so-called beta velocity law,
\begin{equation}
\label{paris}
v(r)=\vinfbet\zav{1-\frac{R_*}{r}}^\beta,
\end{equation}
where {\vinfbet} is the wind terminal velocity, $\beta$ a free parameter
describing the steepness of the velocity law, and $R_*$ the stellar radius.
Thus, for parameterization of the wind velocity law using Eq.~\ref{paris} we
need two parameters, namely $\beta$ and the wind terminal velocity $\vinfbet$.
The popularity of the beta velocity law Eq.~\ref{paris} stems from its
simplicity and from the belief that the velocity profile described by
Eq.~\ref{paris} is close to the real one. Also simple analytical wind solutions
have the form of Eq.~\ref{paris}. For example, in the case of accelerating force
proportional to $\sim 1/r^2$, the solution of the momentum equation
\citep{milne,chandra,rubl} leads to $v(r)=v_{\infty,\beta}\sqrt{1-{R_*}/{r}}$,
implying $\beta=0.5$ in Eq.~\ref{paris}. Also the velocity law of a classical
\cite{cak} line-driven wind model can be approximated with $\beta=0.5$
\citep[c.f.,][]{negra}. Motivated by this, observational studies widely use the
Eq.~\ref{paris} and usually consider $\beta$ as a free parameter, yielding a
typical value of $\beta=0.7-1$ for O star winds \citep[e.g.,][]{pulmoc}.
However, for some stars (e.g., early B supergiants), higher $\beta$ up to 3 may
be determined \citep[e.g.][]{vysbeta}.

Several variants of the simple formula Eq.~\ref{paris} can be found in the
literature and generally aim to improve the fits of wind line profiles. For
example, \citet{snadjo} introduced the velocity law that also encompasses the
wind part close to the hydrostatic photosphere. Subsequently, \citet{hilmibet}
introduced a second component in the beta law, which provides better fits to the
blue edges of P Cygni line profiles. This was also supported by \cite{graham},
who found that different values of $\beta$ describe the wind velocity law in
different regions. The failure of the simple $\beta$-velocity law to describe
the wind velocity law was also mentioned by \cite{goc}.

Although the original beta velocity law was derived theoretically, there is no
reason to believe that it provides the best approximation to the real situation.
It is well known from hydrodynamical simulations \citep{felpulpal,opsim} that
the wind velocity field is variable with time, and consequently the beta law may
at most describe mean flow. But even in stationary models, the wind velocity law
may deviate from Eq.~\ref{paris}, at least because the wind is driven by
different ions close to the star and at large radii \citep{vikolamet,nlteii}.

It would be highly desirable to develop
formulae
that are able to describe the
velocity law in a more accurate form than Eq.~\ref{paris}. It is not clear how
to develop a more general formula based on Eq.~\ref{paris}. Luckily, the
problem of functional fitting is 
fully
settled in mathematics, where the concept
of orthogonal functions was introduced. Consequently, we propose to use the set
of polynomials that may form an orthogonal basis in a given functional space,
e.g.,
\begin{subequations}
\label{vulkan}
\begin{equation}
\label{tuvok}
v(r)=\sum_{i=1}^{n} v_i \zav{1-\frac{R_*}{r}}^i,
\end{equation}
from which the terminal velocity follows as $\vinfty=\sum_{i=1}^{n} v_i$.
Equation~\ref{tuvok} can be {\em equivalently} rewritten in terms of Legendre
polynomials. This guarantees that the derived coefficients decrease with
increasing order $i$ and that the coefficients do not significantly change with
increasing $n$ \citep[e.g.,][]{mik}. In this case instead of Eq.~\ref{tuvok},
the parameterizations is
\begin{equation}
\label{spok}
v(r)=\sum_{i=0}^{n} \tilde v_i \tilde P_i(x),
\qquad\text{with}\;x=1-\frac{R_*}{r},
\end{equation}
\end{subequations}
and the terminal velocity $\vinfty=\sum_{i=1}^{n} \tilde v_i$. 
To ensure the orthogonality in the interval $[0,1]$ we use shifted Legendre
polynomials here. The first four are \citep[e.g.,][]{koo}
\begin{align}
\label{belana}
\nonumber
\tilde P_0 &= 1, & \tilde P_1 &= 2x - 1,\\*
\tilde P_2 &= 6x^2 - 6x + 1\text{, and} & \tilde P_3 &= 20x^3 - 30x^2 + 12x - 1.
\end{align}
The relations between coefficients $\tilde v_i$ and $v_i$ follow from
Eqs.~\ref{vulkan}
%
and \ref{belana}
\begin{align}
\nonumber
\tilde v_0&=\frac{1}{2}v_1+\frac{1}{3}v_2+\frac{1}{4}v_3, &
\tilde v_1&=\frac{1}{2}v_1+\frac{1}{2}v_2+\frac{9}{20}v_3,\\*
\tilde v_2&=\frac{1}{6}v_2+\frac{1}{4}v_3, & \tilde v_3&=\frac{1}{20}v_3.
\end{align}
Only three of these ($\tilde v_i$) coefficients are independent.

\begin{table*}[t]
\caption{Best-fit parameters to the radial velocity from hydrodynamic models
\citep{cmf1} for selected stars and individual velocity parameterizations.}
\label{sedma}
\begin{center}
\begin{tabular}{llrrrcrccrr}
\hline
\hline
Star  & HD number&&&&\multicolumn{2}{c}{$\beta$ law Eq.~\ref{paris}} &
\multicolumn{4}{c}{Eq.~\ref{spok} for $n=3$}\\
&& $R_*$ & $M$ & $T_\text{eff}$& $v_{\infty,\beta}$ & $\beta$ & $\tilde v_0$ & $\tilde v_1$ &
\multicolumn{1}{c}{$\tilde v_2$} & \multicolumn{1}{c}{$\tilde v_3$}\\
&&$[{R}_\odot]$ & $[{M}_\odot$] & [K] & [\kms]&  & \multicolumn{4}{c}{[\kms]} \\
\hline
\multicolumn{11}{c}{O stars}\\
\hline
$\xi$ Per   &  \object{HD 24912} & $14.0$ & $36$ & $35\,000$& 2130 & 0.70 &  1270 &   980 &  $-220$  &   60 \\
$\iota$ Ori &  \object{HD 37043} & $21.6$ & $41$ & $31\,400$& 2170 & 0.83 &  1190 &  1040 &  $-120$  &   30 \\
15 Mon      &  \object{HD 47839} &  $9.9$ & $32$ & $37\,500$& 3240 & 0.90 &  1700 &  1590 &  $ -90$  &   20 \\
            &  \object{HD 54662} & $11.9$ & $38$ & $38\,600$& 2200 & 0.79 &  1230 &  1070 &  $-180$  &  $-10$ \\
            &  \object{HD 93204} & $11.9$ & $41$ & $40\,000$& 2210 & 0.79 &  1240 &  1070 &  $-170$  &    0 \\
$\zeta$ Oph & \object{HD 149757} &  $8.9$ & $21$ & $32\,000$& 1940 & 0.85 &  1060 &   930 &  $ -50$  &   70 \\
68 Cyg      & \object{HD 203064} & $15.7$ & $38$ & $34\,500$& 1970 & 0.76 &  1120 &   920 &  $-150$  &   50 \\
19 Cep      & \object{HD 209975} & $22.9$ & $47$ & $32\,000$& 2090 & 0.82 &  1160 &  1010 &  $-120$  &   30 \\
\hline
\multicolumn{11}{c}{Central stars of planetary nebulae}\\
\hline
\object{NGC 2392}  && 1.5 &0.41 &40\,000&  540 & 0.86 &   290 &   270 &   $-50$  &  $-20$ \\
\object{NGC 3242}  && 0.3 &0.53 &75\,000& 2070 & 0.68 &  1260 &   920 &  $-290$  &   60 \\
\object{IC 4637}   && 0.8 &0.87 &55\,000& 1350 & 0.89 &   710 &   680 &   $-70$  &  $-30$ \\
\object{IC 4593}   && 2.2 &1.11 &40\,000&  730 & 0.88 &   390 &   360 &   $-60$  &  $-30$ \\
\object{IC 418}    && 2.7 &1.33 &39\,000&  800 & 0.87 &   420 &   380 &   $-70$  &  $-20$ \\
\object{Tc 1}      && 3.0 &1.37 &35\,000&  910 & 0.95 &   460 &   440 &   $-30$  &  $-10$ \\
\object{NGC 6826}  && 2.2 &1.40 &44\,000&  880 & 0.88 &   470 &   430 &   $-60$  &  $-30$ \\
\hline
\end{tabular}
\end{center}
\tablefoot{The stellar parameters (radius $R_*$, mass $M$, and the effective
temperature $T_\text{eff}$) are taken from \citet{rep,upice,btpau}, and
\citet{martclump}.}
\end{table*}

\begin{figure}[t]
\centering
\resizebox{0.99\hsize}{!}{\includegraphics{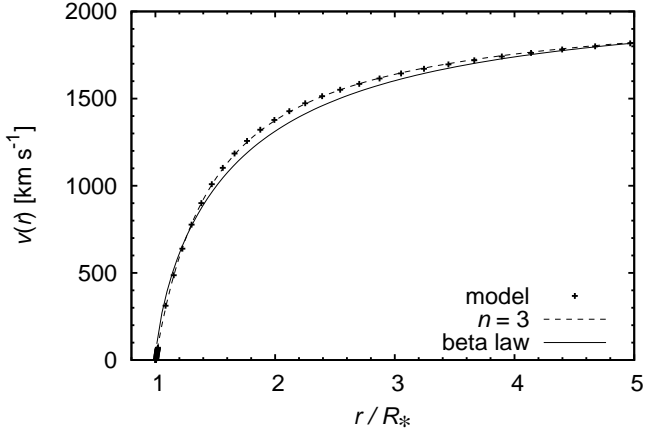}}
\caption{Comparison of the fit to the calculated hydrodynamical model radial
velocity of \object{HD~24912} (crosses) by the beta velocity law (solid line),
and by the polynomial fit Eq.~\ref{spok} for $n=3$.}
\label{24912}
\end{figure}

\begin{figure}[t]
\centering
\resizebox{0.85\hsize}{!}{\includegraphics{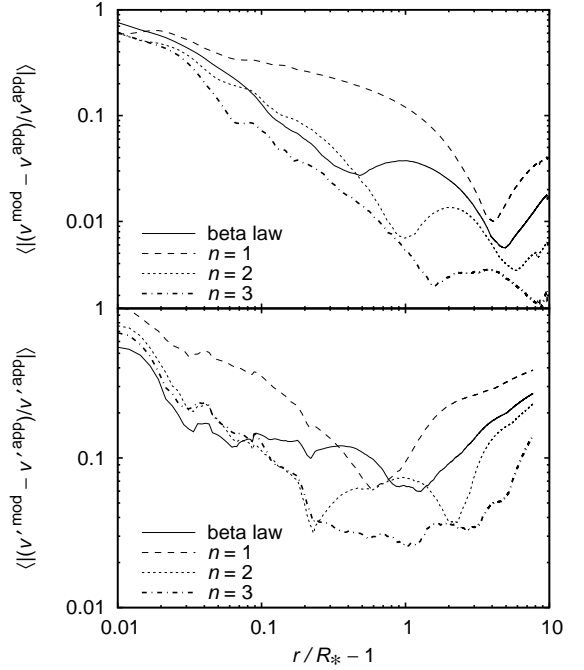}}
\caption{The mean relative difference (see text) between the velocity law
(upper panel) and its derivative $v^\prime = {\text d}v/{\text d} r$
(bottom panel) taken from hydrodynamical models and its approximations
averaged over stars from Table~\ref{sedma} as a function of radius.}
\label{harry}
\end{figure}

To demonstrate the convenience of the formulas Eq.~\ref{vulkan} we fit the
radial velocity structure obtained from hydrodynamical NLTE wind models with CMF
line force of \citet[see Fig.~\ref{24912}]{cmf1}. The best-fit parameters of the
beta velocity law and polynomial fit Eq.~\ref{spok} with $n=3$ for individual
stars are given in Table~\ref{sedma}. To test the global accuracy of individual
orders of the velocity approximation Eq.~\ref{vulkan}, we plot the relative
difference between the velocity from the numerical models $\vmod$ and its
approximations $\vapp$ (Eq.~\ref{spok} for $n=1,2,3$ and Eq.~\ref{paris})
averaged over individual stars in Table~\ref{sedma},
$\langle|(\vmod-\vapp)/\vapp|\rangle$, as a function of radius in
Fig.~\ref{harry}.

Already the first term in Eq.~\ref{tuvok}, i.e. the approximation $v(r)=v_1
\zav{1-{R_*}/{r}}$ (in fact Eq.~\ref{paris} for $\beta=1$), provides reasonable
approximation to the wind velocity law suitable for simplified analysis (see
Fig.~\ref{harry}). Including the second term in Eq.~\ref{tuvok} we obtain
better approximation than the usual beta velocity law both for the velocity and
its derivative using the same number of free parameters ($\beta$ and $\vinfbet$
for the beta velocity law, $v_1$ and $v_2$ for Eq.~\ref{tuvok}). The agreement
in the outer parts of the wind ($r/R_*\gtrsim1.1$) can be improved further by
adding the third term in Eq.~\ref{tuvok} (see Fig.~\ref{harry}).

It has not escaped our attention that even relatively high order polynomials do
not provide reasonable fits in the region close to the star, for
$r/R_*\lesssim1.1$. We do not aim to fit the velocity in this region, because
here the wind density approaches the hydrostatic density stratification.
Consequently, it would be better to derive the velocity from the continuity
equation using density from the hydrostatic equilibrium equation instead of a
polynomial.

Many studies fit the observed line profiles of particular stars using
sophisticated radiative transfer calculations \cite[e.g.,][]{pulmoc,hilmibet} by
{\em assuming} the velocity law of type Eq.~\ref{paris} and trying to find
the most suitable value of the parameter $\beta$ in combination with the value
of the terminal velocity \vinfty. Although our formulae Eq.~\ref{vulkan} were
derived to fit the results of numerical hydrodynamical calculations, it might be
interesting to use them to also fit the beta velocity profiles. The test showed
that the polynomial approximation Eq.~\ref{vulkan} with $n=2$ provides very good
fits to beta laws (Eq.~\ref{paris}) both with small $\beta<1$ and large $\beta>1$
and to the combined law of \citet{hilmibet}. Only for large $\beta\gtrsim2.3$ is it
better to increase the polynomial degree ($n=3$) to assure the positivity of
the fit close to the star. The modified beta law
$v(r)=\vinfbet\zav{1-b\frac{R_*}{r}}^\beta$ with additional parameter $b<1$
\citep{pulshumra} is frequently used to improve the fit close to the star. It
can be satisfactorily reproduced by either introducing a fixed low value of $v_0$ in
Eq.~\ref{tuvok} or by a substitution $x=1-bR_*/r$ in Eq.~\ref{spok}. Moreover,
our experiments with fitting the trial P~Cygni line profiles have shown that
the fits using Eq.~\ref{vulkan} and the beta velocity law require a comparable
number of line profile evaluations. A reasonable fit strategy could be to derive
$\vinfty$ from observations and then keep the sum $\sum_{i=1}^{n} v_i=\vinfty$
fixed.

We conclude that the beta velocity law provides reasonable approximation to the
hydrodynamical wind velocity law and its derivative. The approximation can be
improved with polynomial fits Eq.~\ref{tuvok} (or, equivalently
Eq.~\ref{spok}) using the same number of free parameters as in the commonly
used beta velocity law.

\begin{acknowledgements}
We thank Prof.~Achim Feldmeier and Dr.~Joachim Puls for discussions of this
subject. This work was supported by grant GA \v{C}R 205/08/0003. The
Astronomical Institute Ond\v{r}ejov is supported by the project AV0\,Z10030501.
\end{acknowledgements}


\begin{thebibliography}{}
\bibitem[Castor, Abbott \& Klein(1975)]{cak} Castor, J. I., Abbott, D. C.,
	\& Klein, R. I. 1975, ApJ, 195, 157 (CAK)
\bibitem[Chandrasekhar(1934)]{chandra} Chandrasekhar, S.  1934, MNRAS, 94, 522
\bibitem[Crowther et al.(2006)]{vysbeta} Crowther, P. A., Lennon, D. J.,
	\& Walborn, N. R. \citeyear{vysbeta}, A\&A, 446, 279
\bibitem[Feldmeier et al.(1997)]{felpulpal} Feldmeier, A., Puls, J., \&
        Pauldrach, A. W. A. 1997, A\&A, 322, 878
\bibitem[Gayley(2000)]{negra} Gayley, K. G. 2000, ApJ, 529, 1019
\bibitem[Gayley et al.(1995)]{goc} Gayley, K. G., Owocki, S. P., \& Cranmer,
        S. R. 1995, ApJ, 442, 296
\bibitem[Gr\"afener \& Hamann(2005)]{graham} Gr\"afener, G., \& Hamann, W.-R.
        2005, A\&A, 432, 633
\bibitem[Hillier(1988)]{snadjo} Hillier, D. J. 1988, ApJ, 327, 822
\bibitem[Hillier \& Miller(1999)]{hilmibet} Hillier, D. J., \& Miller, D. L.
        1999, ApJ, 519, 354
\bibitem[Koornwinder et al.(2010)]{koo} Koornwinder, T. H., Wong, R. S. C.,
        Koekoek, R., \& Swarttouw, R. F. 2010, in  F. W. J. Olver, D. M. Lozier,
        R. F. Boisvertet et al., NIST Handbook of Mathematical Functions
        (Cambridge: Cambridge University Press)
\bibitem[Krti\v cka(2006)]{nlteii} Krti\v cka, J. 2006, MNRAS, 367, 1282
\bibitem[Krti\v cka \& Kub\'at(2010)]{cmf1} Krti\v cka, J., \& Kub\'at,
        J.  2010, A\&A, 519, A50
\bibitem[Markova et al.(2004)]{upice} Markova, N., Puls, J., Repolust,  T.,
        \&  Markov, H. 2004, A\&A, 413, 693
\bibitem[Martins et al.(2005)]{martclump} Martins, F., Schaerer, D.,
        Hillier, D. J., et al. 2005, A\&A, 441, 735
\bibitem[Mikul\'a\v sek(2007)]{mik} Mikul\'a\v sek, Z. 2007, Odessa Astronomical
        Publications, 20, 138
\bibitem[Milne(1926)]{milne} Milne, E. A. 1926, MNRAS, 86, 459	
\bibitem[Owocki \& Puls(1999)]{opsim} Owocki, S. P., \& Puls, J. 1999, ApJ, 510,
	355
\bibitem[Pauldrach et al.(2004)]{btpau} Pauldrach, A. W. A., Hoffmann, T. L., \&
        M\'endez, R. H. 2004, A\&A, 419, 1111
\bibitem[Puls \& Hummer(1988)]{pulshumra} Puls, J., \& Hummer, D. G. 1988, A\&A, 191,
	87
\bibitem[Puls et al.(1996)]{pulmoc} Puls, J., Kudritzki, R.-P., Herrero, A.,
        et al. 1996, A\&A, 305, 171
\bibitem[Repolust et al.(2004)]{rep} Repolust, T., Puls, J., \& Herrero, A.
        2004, A\&A, 415, 349
\bibitem[Rublev(1965)]{rubl} Rublev, S. V. 1965, AZh, 42, 718
        (Soviet Astronomy, 9, 555)
\bibitem[Vink et al.(2001)]{vikolamet} Vink, J. S., de Koter, A., \&
	Lamers, H. J. G. L. M. 2001, A\&A, 369, 574
\end{thebibliography}
\end{document}